\documentclass[12pt,reqno]{amsart}
\usepackage{eurosym}
\usepackage{amsfonts}
\usepackage{graphicx}
\usepackage{amscd}
\usepackage{amsmath}
\usepackage{epsfig}
\usepackage[usenames]{color}
\usepackage{subfigure}
\usepackage[english]{babel}

\providecommand{\U}[1]{\protect\rule{.1in}{.1in}}
\providecommand{\U}[1]{\protect\rule{.1in}{.1in}}
\allowdisplaybreaks

\newtheorem{lemma}{Lemma}
\newtheorem{proposition}{Proposition}
\theoremstyle{definition}

\theoremstyle{plain}

\newtheorem{example}{Example}

\newtheorem{remark}{Remark}

\numberwithin{equation}{section}

\DeclareMathOperator{\sech}{sech}

 \email{tamer.oraby@utrgv.edu}
\email{allan.g.martinez01@utrgv.edu}
\email{carlos.montes02@utrgv.edu}
\email{kolade.adjibi01@utrgv.edu}
\email{miguel.mascorro01@utrgv.edu}
\email{erwin.suazo@utrgv.edu}
\email{rita.sandoval01@utrgv.edu}

\begin{document}
\title[Exact Solutions of Stochastic Burgers-KdV Equations]{Exact Solutions of
Stochastic Burgers-KdV Equation with Variable Coefficients}
\author{}
\maketitle

\centerline{\scshape Kolade Adjibi, Allan Martinez, Miguel Mascorro}
\medskip %{\footnotesize
%% please put the address of the first author
% \centerline{School of Mathematical and Statistical Sciences, The University of Texas Rio Grande Valley}
%   \centerline{1201 W. University Drive,}
%   \centerline{Edinburg, Texas, 78539, USA}
%} % Do not forget to end the {\footnotesize by the sign }
\centerline{\scshape Carlos Montes, Tamer Oraby, Rita Sandoval, and Erwin Suazo$^*$} \medskip {\footnotesize \ 
\centerline{School of Mathematical and Statistical Sciences, The University
of Texas Rio Grande Valley} \centerline{1201 W. University Drive} %
\centerline{Edinburg, Texas, 78539, USA} }

\begin{abstract}
We will present exact solutions for three variations of stochastic Korteweg de Vries-Burgers (KdV-Burgers) equation featuring variable coefficients. In each variant, white noise exhibits spatial uniformity, and the three categories include additive, multiplicative, and advection noise. Across all cases, the coefficients are time-dependent functions. Our discovery indicates that solving certain deterministic counterparts of KdV-Burgers equations and composing the solution with a solution of stochastic differential equations leads to the exact solution of the stochastic Korteweg de Vries-Burgers (KdV-Burgers) equations.
\end{abstract}

% Enter the first author's name and address:

%\medskip

% please put the address of the second  and third author

%\bigskip
%
%% The name of the associate editor will be entered by an editorial staff
%% "Communicated by the associate editor name" is not needed for special issue.
% \centerline{(Communicated by the associate editor name)}

%The abstract of your paper

%The title of your section 1

\section{Introduction}
The majority of physical and biological systems exhibit nonhomogeneity, often influenced by environmental fluctuations and the existence of nonuniform mediums. Consequently, the nonlinear equations relevant to practical applications typically involve coefficients that vary spatially and/or temporally, along with stochastic terms. Reaction-diffusion equations are crucial in modeling heat diffusion and reaction processes in nonlinear acoustics, biology, chemistry, genetics, and various other research domains. However, like numerous mathematical models representing real-world phenomena, solving this problem explicitly poses a considerable challenge.

The Burgers-Korteweg-de Vries equation (Burgers-KdV) arises from many physical contexts, for example, the propagation of undular wells in shallow water \cite{Benney}, the flow of liquids containing gas bubbles \cite{Wijngaarden}, the propagation of waves in an elastic tube filled with a viscous fluid \cite{Johnson}, and weakly nonlinear plasma waves with certain dissipative effects \cite{Grad, Hu}. It is also used as a non-linear model in crystal lattice theory, non-linear circuit theory, and turbulence \cite{Gao,  Wadati }.

The goal of this paper is to introduce exact solutions for stochastic Korteweg-de Vries-Burgers equations with variable coefficients with space-uniform white noise. We consider the following three different
stochastic KdV-Burgers equations (additive, advection, and multiplicative noise) with a space-uniform white noise of the
form 
\begin{equation}
du=(\delta (t)\partial _{zzz}u+\beta (t)u\partial _{z}u+\mu (t)\partial
_{zz}u+\alpha (t)\partial _{z}u+\gamma (t)u)dt+\sigma (t)\partial _{z}u
dW_{t}  \label{eq1}
\end{equation}%
\begin{equation}
du=(\delta (t)\partial _{zzz}u+\beta (t)u\partial _{z}u+\mu (t)\partial
_{zz}u+\alpha (t)\partial _{z}u+\gamma (t)u)dt+\sigma (t) dW_{t}
\label{eq1_10}
\end{equation}%
for $t\in \lbrack t_{0},T]$ and $z\in \mathbb{R}$ with $u(0,z)=\phi (z)$ for 
$z\in \mathbb{R}$. We also consider a linear PDE in the KdV form of
\begin{equation}
du=(\delta (t)\partial _{zzz}u+\mu (t)\partial
_{zz}u+\alpha (t)\partial _{z}u+\gamma (t)u)dt+\sigma (t) u dW_{t}
\label{eq1_11}
\end{equation}%
for $t\in \lbrack t_{0},T]$ and $z\in \mathbb{R}$ with $u(0,z)=\phi (z)$ for 
$z\in \mathbb{R}$.

%In the underlying SBE, we consider here noise added to the coefficient of the velocity field and as a forcing term.
%
%In case $A\equiv c$ (constant), $B\equiv 1$, $C\equiv 0$, and $D\equiv 0$, Cole-Hopf transformation $u=2c \partial_x \log(v)$ results in $v$ solving the heat equation given by
%\begin{equation}\label{eqheat}
%\partial_t v=c\partial_{zz}v, \quad v(0,z)=\exp(\frac{1}{2c}\int_0^z \phi(y) dy)
%\end{equation}

This paper is organized as follows: In Section 2, we recall two lemmas that provide details and properties in solving SDEs and execute numerical simulations. In Section 3, We present exact solutions for stochastic Korteweg de Vries-Burgers (KdV-Burgers) equations \eqref{eq1}-\eqref{eq1_11} featuring variable coefficients. In each variant, white noise exhibits spatial uniformity, and the three categories include additive, multiplicative, and advection noise.
Across all cases, the coefficients are time-dependent functions. Our discovery indicates that solving certain deterministic counterparts of KdV-Burgers equations and composing the solution with a solution of stochastic differential equations leads to the exact solution of the stochastic Korteweg de
Vries-Burgers (KdV-Burgers) equations. We provide several examples.

\section{Preliminaries}

Consider the probability space $(\Omega,\mathcal{F},\mathbf{P})$ for which
the Brownian motion $\{W_t,t\geq 0\}$ is defined and $E(W_sW_t)=\min(s,t)$
for all $s,t\geq 0$. Also consider the filtration $\mathcal{F}%
_t:=\sigma(W_s:s \leq t)$ being the smallest $\sigma-$algebra to which $W_s$
is measurable for $s \leq t$.

%\subsection{It\^o Calculus and Stochastic Differential Equations}
Then consider the stochastic differential equation (SDE) with variable
coefficients \cite{Kloeden1992}%
\begin{equation}  \label{SDE1}
dX_{t}=\alpha(t,X_t)dt+ \beta(t,X_t)dW_{t},
\end{equation}
with initial state $X_{t_0}$ and for $t\in [t_0,T]$. The SDE in \eqref{SDE1}
has a general solution given by 
\begin{equation*}
X_t= X_{t_0}+\int_{t_0}^t \alpha(s,X_s) ds+\int_{t_0}^t \beta(s,X_s) dW_s
\end{equation*}
for $t\leq T$. If $\alpha(t):=\alpha(t,X_t)$ and $\beta(t):=\beta(t,X_t)$,
then equation \eqref{SDE1} has a general solution given by 
\begin{equation*}
X_t= X_{t_0}+\int_{t_0}^t \alpha(s) ds+\int_{t_0}^t \beta(s) dW_s
\end{equation*}
for $t\leq T$. The process $\{W_t;t \geq 0\}$ is a Wiener process with
respect to filtration $\{\mathcal{F}_t;t\geq 0\}$. The initial state $%
X_{t_0}$ is $\mathcal{F}_{t_0}$ and the functions $\alpha(t)$ and $\beta(t)$
are Lebesgue measurable and bounded on $[t_0,T]$. The latter implies both
the global Lipschitz and linearity growth conditions required to ensure the
existence and (pathwise) uniqueness of a strong solution to \eqref{SDE1}, 
\cite{Kloeden1992}.

Let $X_t$ and $Y_t$ be any two diffusion processes such as those defined by the
solution of equation \eqref{SDE1}. If $F(x,y)$ is a differentiable function
that works as a transformation for two processes $X_t$ and $Y_t$, then the
general bi-variate It\^o formula \cite{Kloeden1992} gives 
\begin{eqnarray}  \label{SDE2}
dF(X_{t},Y_t)&=&\partial_x F(X_t,Y_t)dX_t+ \partial_y
F(X_t,Y_t)dY_{t}+\frac12 \partial_{xx} F(X_t,Y_t) (dX_t)^2 \\
&&+ \frac12 \partial_{yy} F(X_t,Y_t) (dY_{t})^2+\partial_{xy} F(X_t,Y_t)
dX_t dY_t.  \notag
\end{eqnarray}
$F(x,y)$ is a differentiable function. 

%\subsection{Burgers Equations with Variable Coefficients}
The following two lemmas are crucial to identify the solutions of the SDEs, they were introduced previously in \cite{Oraby}. These two lemmas are fundamental for our simulations. We use the lemmas to simulate the processes with $X_0=x$ and then compose the exact solutions with the simulations based on Lemma 3.

\begin{lemma}
\label{lemma0}

\begin{enumerate}
\item The stochastic process $X_{t}$ solving 
\begin{equation*}
dX_{t}=C(t)dt+ E(t)dW_{t}
\end{equation*}
with $X_{t_0}\sim N(x_{t_0},\sigma^2_0)$ independent of $W_t$, is a
non-stationary Gaussian process with mean $x_{t_0}+\int_{t_0}^t C(s) ds$ and
variance $\sigma^2(X_t)=\sigma^2_0+\int_{t_0}^t E^2(s) ds$.

\item The covariance of the two processes $X_t$ and $W_t$ is 
\begin{equation*}
\sigma(X_t,W_t)=\int_{t_0}^t E(s) d{s}.
\end{equation*}

\item Moreover, 
\begin{equation*}
[X_t|W_t=w]\sim N \left(x_{t_0}+\int_{t_0}^t C(s) ds+\frac{w\int_{t_0}^t
E(s) d{s}}{t},V^2_1(t) \right)
\end{equation*}
where $V^2_1(t)=\sigma^2_0+\int_{t_0}^t E^2(s) ds-\frac{(\int_{t_0}^t E(s) d{%
s})^2}{t}$.
\end{enumerate}
\end{lemma}

%
%\begin{proof}
%See \cite[Proposition 5.6.1]{Calin2015} for the first part and \cite[p.86]%
%{Kloeden1992} for the second part. The third point follows from the fact
%that if 
%\begin{equation*}
%\begin{pmatrix}
%X_1 \\ 
%X_2%
%\end{pmatrix}
%\sim N(%
%\begin{pmatrix}
%\mu_1 \\ 
%\mu_2%
%\end{pmatrix}%
%,%
%\begin{pmatrix}
%\sigma_1^2 & \sigma_{1,2} \\ 
%\sigma_{1,2} & \sigma_2^2%
%\end{pmatrix}%
%)
%\end{equation*}
%then the conditional distribution of $X_1$ given $X_2$ is given by 
%\begin{equation*}
%[X_1|X_2]\sim N \left(\mu_1+\dfrac{\sigma_{1,2}}{\sigma_2^2}(X_2-\mu_2),
%\sigma_1^2-\frac{\sigma_{1,2}}{\sigma_2^2}\right),
%\end{equation*}
%see \cite{casella2002statistical}.
%\end{proof}
%

\begin{lemma}
\label{lemma1}

\begin{enumerate}
\item The position random process $Z_t:=z+\int_{t_0}^t \bar{B}(s) K(s)dW_{s}$
solves the Langevin-type second-order SDE 
\begin{equation*}
\ddot{Z}_{t}=\dfrac{B^{\prime }(t)}{B(t)} \dot{Z}_t+B(t)K(t)\dot{W}_t, \,\,
t\in [t_0,T]
\end{equation*}
with initial state $Z_{t_0}=z$, where $\bar{B}(s)=\int_{s}^t B(r)dr$ for $%
t>s $.

\item The process $Z_t$ is a nonstationary Gaussian process with mean $z$
and variance $\sigma^2(Z_t)=\int_{t_0}^t (\bar{B}(s) K(s))^2 ds$.

\item Meanwhile, 
\begin{equation*}
\dot{Z}_t:= B(t) \left(\int_{t_0}^t K(r)dW_{r}\right).
\end{equation*}

\item The process $\dot{Z}_t$ is a non-stationary Gaussian process with mean
zero and variance $\sigma^2(\dot{Z}_t)=(B(t))^2\int_{t_0}^t ( K(s))^2 ds$.

\item The covariance of $Z_t$ and $W_t$ is 
\begin{equation*}
\sigma(Z_t,W_t)=\int_{t_0}^t \bar{B}(s) K(s)d{s},
\end{equation*}
\begin{equation*}
\sigma(\dot{Z}_t,W_t)=B(t) \left(\int_{t_0}^t K(s)ds\right).
\end{equation*}

\item The conditional distributions are given by
\end{enumerate}

\begin{equation*}
[Z_t|W_t=w]\sim N \left(z+w\frac{\int_{t_0}^t \bar{B}(s) K(s)d{s}}{t},
\int_{t_0}^t (\bar{B}(s) K(s))^2 ds-\frac{(\int_{t_0}^t \bar{B}(s) K(s)d{s}%
)^2}{t}\right)
\end{equation*}
and 
\begin{equation*}
[\dot{Z}_t|W_t=w]\sim N \left(\frac{w B(t) \left(\int_{t_0}^t K(s)ds\right)}{%
t}, (B(t))^2 \left[\int_{t_0}^t ( K(s))^2 ds-\frac{ \left(\int_{t_0}^t
K(s)ds\right)^2}{t}\right]\right).
\end{equation*}
\end{lemma}
\section{Stochastic Burgers-KdV Equation}
In this Section, through the use of Ito calculus and interesting transformations. We present exact solutions for stochastic Korteweg de Vries-Burgers (KdV-Burgers) equations \eqref{eq1}-\eqref{eq1_11} featuring variable coefficients. As the following lemma shows, solving certain deterministic counterparts of KdV-Burgers equations and composing the solution with a solution of stochastic differential equations leads to the exact solution of the stochastic Korteweg de
Vries-Burgers (KdV-Burgers) equations. We provide several examples.

\begin{lemma}
\label{thm1} Let $\alpha ,\beta ,\gamma ,\delta ,\mu ,\sigma \in \mathcal{C}%
^{b}\left( [t_{0},T]\right) $ be bounded functions on $[t_{0},T]$. Assume
that $\beta (t)>0$ for all $t\in \lbrack t_{0},T]$. Then we have
\end{lemma}

\begin{enumerate}
\item The stochastic Burgers-KdV equation \eqref{eq1} has a solution $u(t,z)=U(t,X_{t})$, where $%
U(t,x)$ is the solution of 
\begin{equation}
\partial _{t}U=\delta (t)\partial _{xxx}U+(\mu (t)-\frac{1}{2}\sigma
^{2}(t))\partial _{xx}U+\beta (t)U\partial _{x}U+\gamma (t)u,\quad
U(0,x)=\phi (x)  \label{eq01}
\end{equation}%
and $X_{t}$ is the solution of 
\begin{equation}
dX_{t}=\alpha (t)dt+\sigma (t)dW_{t}  \label{SDE2}
\end{equation}%
with initial state $X_{t_{0}}=z$ and for $t\in \lbrack t_{0},T]$.

\item The stochastic Burgers-KdV equation with the initial value problem %
\eqref{eq1_10} has a solution 
\begin{equation*}
u(t,z)=R(t)\left( V(t,Z_{t})+\dfrac{1}{\mathfrak{B}(t)}\dot{Z}_{t}\right)
\end{equation*}%
where $V(t,x)$ is the solution of 
\begin{equation}
\partial _{t}V=\delta (t)\partial _{xxx}V+\mu (t)\partial _{xx}V+\mathfrak{B}%
(t)V\partial _{x}V+\alpha (t)\partial _{x}V,\quad V(0,x)=\phi (x),
\label{eq010}
\end{equation}%
and $Z_{t}$ is the solution of a second-order stochastic differential
equation 
\begin{equation}
\ddot{Z}_{t}=\dfrac{\mathfrak{B}^{\prime }(t)}{\mathfrak{B}(t)}\dot{Z}_{t}+%
\dfrac{\mathfrak{B}(t)\sigma (t)}{R(t)}\dot{W}_{t},  \label{SDE2_2}
\end{equation}%
with initial state $Z_{t_{0}}=z$ and for $t\in \lbrack t_{0},T]$. Also, $%
R(t)=\exp (\int_{t_{0}}^{t}\gamma (s)ds)$ and $\mathfrak{B}(t)=\beta (t)R(t)$%
.
\end{enumerate}

\begin{proof}
For (1), apply It\^{o}'s formula to the $X_{t}$ solution of \eqref{SDE2}
with the transformation $U(t,x)$ that solves the deterministic KdV-Burgers
equation \eqref{eq01} 
\begin{equation}
dU(t,X_{t})=f(t,X_{t})dt+g(t,X_{t})dW_{t},  \label{SDE21}
\end{equation}%
where 
\begin{equation*}
f(t,x)=\partial _{t}U(t,x)+\alpha (t)\partial _{x}U(t,x)+\frac{1}{2}\sigma
^{2}(t)\partial _{xx}U(t,x).
\end{equation*}%
Note that, 
\begin{equation*}
\partial _{t}U(t,x)=\delta (t)\partial _{xxx}U+(\mu (t)-\frac{1}{2}\sigma
^{2}(t))\partial _{xx}U(t,x)+\beta (t)U(t,x)\partial _{x}U(t,x)+\gamma
(t)U(t,x).
\end{equation*}%
Therefore, 
\begin{equation*}
f(t,x)=\delta (t)\partial _{xxx}U+\mu (t)\partial _{xx}U(t,x)+\beta
(t)U(t,x)\partial _{x}U(t,x)+\alpha (t)\partial _{x}U(t,x)+\gamma (t)U(t,x).
\end{equation*}%
Observe that 
\begin{equation*}
g(t,x)=\sigma (t)\partial _{x}U(t,x),
\end{equation*}%
which proves (1).

To prove (2). Let's take $u(t,z)=R(t)\,L(t,z)$. By the bi-variate general of
It\^{o}'s formula we get 
\begin{equation}
du(t,z)=R^{\prime }(t)L(t,z)dt+R(t)dL(t,z).  \label{du}
\end{equation}

In order to prove (\ref{eq1_10}) we need to show 
\begin{equation}
dL(t,z)=(\delta (t)\partial _{zzz}L(t,z)+\mu (t)\partial _{zz}L(t,z)+%
\mathfrak{B}(t)L(t,z)\partial _{z}L(t,z)+\alpha (t)\partial _{z}L(t,z))dt+%
\dfrac{\sigma (t)}{R(t)}d{W}_{t}.  \label{dL}
\end{equation}%
Before proving (\ref{dL}), let's show how this statement will finish the
proof of (\ref{eq1_10}). Recalling that $R(t)=\exp (\int_{t_{0}}^{t}\gamma
(s)ds)$ and using \eqref{dL}, from \eqref{du} we obtain

\begin{eqnarray*}
du &=&R(t)(\delta (t)\partial _{zzz}L(t,z)+\mu (t)\partial _{zz}L(t,z)+%
\mathfrak{B}(t)L(t,z)\partial _{z}L(t,z)+\alpha (t)\partial _{z}L(t,z))dt \\
&&+\gamma (t)R(t)L(t,z)dt+\sigma (t)d{W}_{t}
\end{eqnarray*}

Hence, we obtain (\ref{du}) as we wanted.

Let's proceed to prove \eqref{dL}, equation \eqref{SDE2_2} can be written as
a system of equations of the following form 
\begin{eqnarray}
d{N}_{t} &=&\dfrac{\mathfrak{B}^{\prime }(t)}{\mathfrak{B}(t)}N_{t}dt+\dfrac{%
\mathfrak{B}(t)\sigma (t)}{R(t)}d{W}_{t}, \\
dZ_{t} &=&N_{t}dt.
\end{eqnarray}%
Let $V(t,x)$ be the solution of 
\begin{equation*}
\partial _{t}V=\delta (t)\partial _{xxx}V+\mu (t)\partial _{xx}V+\mathfrak{B}%
(t)V\partial _{x}V+\alpha (t)\partial _{x}V,\quad V(0,x)=\phi (x)
\end{equation*}%
Using the bivariate general It\^{o} formula for $V(t,Z_{t})$ we obtain 
\begin{equation*}
dV(t,Z_{t})=\partial _{t}V(t,Z_{t})dt+\partial _{x}V(t,Z_{t})dZ_{t},
\end{equation*}%
since $(dt)^{2}=0$, $dtdZ_{t}=N_{t}(dt)^{2}=0$, and $%
(dZ_{t})^{2}=(N_{t})^{2}(dt)^{2}=0$. Thus, 
\begin{eqnarray}
dV(t,Z_{t}) &=&\left( \delta (t)\partial _{xxx}V(t,Z_{t})+\mu (t)\partial
_{xx}V(t,Z_{t})+\mathfrak{B}(t)V(t,Z_{t})\partial _{x}V(t,Z_{t})+\alpha
(t)\partial _{x}V(t,Z_{t})\right) dt  \notag \\
&&+\partial _{x}V(t,Z_{t})N_{t}dt  \label{eqninter}
\end{eqnarray}%
If we define $L(t,z)=V(t,Z_{t})+\dfrac{1}{\mathfrak{B}(t)}\dot{Z}_{t}$, we
can rewrite the right-hand side of equation \eqref{eqninter} as%
\begin{equation*}
(\delta (t)\partial _{zzz}L(t,z)+\mu (t)\partial _{zz}L(t,z)+\mathfrak{B}%
(t)(L(t,z)-\dfrac{1}{\mathfrak{B}(t)}\dot{Z}_{t})\partial _{z}L(t,z)+\alpha
(t)\partial _{z}L(t,z)+\partial _{z}L(t,z)N_{t})dt
\end{equation*}%
\begin{equation*}
=(\delta (t)\partial _{zzz}L(t,z)+\mu (t)\partial _{zz}L(t,z)+\mathfrak{B}%
(t)L(t,z)\partial _{z}L(t,z)+\alpha (t)\partial _{z}L(t,z))dt.
\end{equation*}%
It follows that, 
\begin{equation*}
dL(t,z)=(\delta (t)\partial _{zzz}L(t,z)+\mu (t)\partial _{zz}L(t,z)+%
\mathfrak{B}(t)L(t,z)\partial _{z}L(t,z)+\alpha (t)\partial
_{z}L(t,z))dt+d\left( \dfrac{1}{\mathfrak{B}(t)}\dot{Z}_{t}\right) .
\end{equation*}%
Applying bivariate general It\^{o} formula It\^{o}'s formula once again, we
obtain 
\begin{equation*}
d\left( \dfrac{1}{\mathfrak{B}(t)}N_{t}\right) =\dfrac{-\mathfrak{B}^{\prime
}(t)}{(\mathfrak{B}(t))^{2}}N_{t}dt+\dfrac{1}{\mathfrak{B}(t)}dN_{t}
\end{equation*}%
since $(dt)^{2}=0$, $dtdN_{t}=0$, and $\dfrac{d^{2}x/\mathfrak{B}(t)}{dx^{2}}%
=0$. Therefore, we have 
\begin{equation*}
d\left( \dfrac{1}{\mathfrak{B}(t)}\dot{Z}_{t}\right) =\dfrac{-\mathfrak{B}%
^{\prime }(t)}{(\mathfrak{B}(t))^{2}}N_{t}dt+\dfrac{1}{\mathfrak{B}(t)}%
\left( \dfrac{\mathfrak{B}^{\prime }(t)}{\mathfrak{B}(t)}N_{t}dt+\dfrac{%
\mathfrak{B}(t)\sigma (t)}{R(t)}d{W}_{t}\right) =\dfrac{\sigma (t)}{R(t)}d{W}%
_{t}.
\end{equation*}%
Hence, we obtain \eqref{dL} as we wanted.
\end{proof}

%
%\begin{proof}
%Since $Z_t:=z+\int_{t_0}^t B(s) \left(\int_{t_0}^s K(r)dW_{r}\right)ds$, by
%the stochastic Fubini's theorem \cite[Theorem 10.3.15]{Calin2015}, part (1)
%follows from the equality 
%\begin{equation*}
%d\left(\dfrac{1}{B(t)} \dot{Z}_t\right)=K(t) d{W}_t
%\end{equation*}
%shown in the previous proof. Part (2) then follows from \cite[Proposition
%5.6.1]{Calin2015}.
%
%Since $\int_{t_0}^s K(r)dW_{r}$ has a continuous sample path (a.s.) \cite[%
%Theorem 3.2.6]{Kloeden1992}, then 
%\begin{equation*}
%\dot{Z}_t:= B(t) \left(\int_{t_0}^t K(r)dW_{r}\right).
%\end{equation*}
%Thus part (3) follows, part (4) follows from \cite[Proposition 5.6.1]%
%{Calin2015}.
%
%The rest of the parts of the lemma follow from the same methods used in
%Lemma \ref{lemma0}.
%\end{proof}

\begin{remark}
\bigskip If $\delta ,$ $\mu $ and $\beta $ are constants, then the
KdV-Burgers equation 
\begin{equation}
\partial _{t}U=\delta \partial _{xxx}U+\mu U_{xx}+\beta U\partial _{x}U,\quad
\end{equation}

has the following explicit solution%
\begin{equation}
U(t,x)=\dfrac{3\mu ^{2}}{25\beta \delta }\sech^{2}\left( \frac{\mu }{10\delta 
}x-\frac{6\mu ^{3}}{250\delta ^{2}}t\right) -\frac{6\mu ^{2}}{25\beta \delta 
}\tanh \left( \frac{\mu }{10\delta }x-\frac{6\mu ^{3}}{250\delta ^{2}}%
t\right) +\frac{6\mu ^{2}}{25\beta \delta }.
\end{equation}
\end{remark}
The following proposition will provide solutions for linear KdV-type equations.
\begin{proposition}
\bigskip Let us consider the stochastic process
\begin{equation}
X_{t}(x)=x+\frac{1}{2}\int_0^t\sigma ^{2}(s)ds-\int_0^t\sigma(s) W_{s}  \label{Stochastic}
\end{equation}
for $x\in \mathbb{R}$, and the equation%
\begin{equation}
du(t,x)=f(u(t,x))dt+\sigma(t) u(t,x) dW_t  \label{ComposeEquation}
\end{equation}
and $f$ is linear function, the equation \eqref{ComposeEquation} can be reduced to 
\begin{equation*}
dv(t,x)=f(v(t,x))dt
\end{equation*}
through the transformation%
\begin{equation*}
v(t,x)=u(t,x)e^{X_t(x)}.
\end{equation*}

Proof: By It\^o's formula we see that%
\begin{equation}
de^{X_{t}}=e^{X_{t}}(\sigma ^{2}(t) dt-\sigma(t) dW_t).  \label{dE}
\end{equation}
And by product rule $dv=ude^{X_{t}}+e^{X_{t}}du+de^{X_{t}}du$ and after
replacing \eqref{ComposeEquation}, \eqref{dE} and \eqref{Stochastic}, we
obtain 
\begin{eqnarray*}
dv &=&ue^{X_{t}}\left( \sigma ^{2}dt-\sigma dW_t\right) +e^{X_{t}}\left( f(u)dt+\sigma
udW_t\right)  \\
&&+\left( f(u)dt+\sigma udW_t\right)e^{X_t} \left( \sigma ^{2}dt-\sigma dW_{t}\right).
\end{eqnarray*}

Finally, using standard It\^o calculus rules and simplifying, we obtain
\begin{equation*}
dv=f(v)dt.
\end{equation*}
\end{proposition}

The following formula will be useful for the following examples, see Chapter
7 by Calin:%
\begin{equation*}
\int_{a}^{b}f(t)dW_{t}=f(t)W_{t}|_{a}^{b}-\int_{a}^{b}f^{\prime }(t)W_{t}dt.
\end{equation*}
%\end{remark}
Next, we provide several examples.
\begin{example}
Consider the stochastic KdV-Burgers equation 
\begin{equation}
du=(\delta \partial _{zzz}u+\beta u\partial _{z}u+\mu \partial _{zz}u+\alpha
(t)\partial _{z}u)dt+\sigma \partial _{z}u\,dW_{t},
\label{KdV Burgers general}
\end{equation}%
where $\delta ,$ $\beta ,\sigma $ and $\mu $ are real constants$.$ By Lemma
1 part (1), equation \eqref{KdV Burgers general} has a solution $%
u(t,z)=U(t,X_{t})$, such that $U(t,x)$ is the solution of 
\begin{equation}
\partial _{t}U=\delta \partial _{xxx}U+\left( \mu -\frac{\sigma ^{2}}{2}%
\right) U_{xx}+\beta U\partial _{x}U.\quad
\end{equation}

In particular, the KdV-Burgers equation ($\delta =\beta =\mu =\alpha =\sigma
=1$) 
\begin{equation}
du=(\partial _{zzz}u+u\partial _{z}u+\partial _{zz}u+\partial
_{z}u)dt+\partial _{z}u\,dW_{t},  \label{KdVBurgers Example 1}
\end{equation}

\bigskip has a solution $u(t,z)=U(t,X_{t})$, such that $U(t,x)$ is given by%
\begin{equation}
U(t,x)=\dfrac{3}{25}\sech^{2}\left( \frac{1}{10}x-\frac{6}{250}t\right) -%
\frac{6}{25}\tanh \left( \frac{1}{10}x-\frac{6}{250}t\right) +\frac{6}{25}
\end{equation}

\bigskip and $X_{t}$ $=z+t+W_{t}$ is the solution of ($\alpha =1$ and $%
\sigma =1$) 
\begin{equation}
dX_{t}=dt+dW_{t},
\end{equation}%
with initial state $X_{0}=z$ and for $t\in \lbrack 0,1]$.

Finally, the explicit solution of \eqref{KdVBurgers Example 1} is given by 
\begin{equation}\label{solnex01}
u(t,z)=U(t,X_{t})=\dfrac{3}{25}\sech^{2}\left( \frac{z+t+W_{t}}{10}-\frac{6t}{%
250}\right) -\frac{6}{25}\tanh \left( \frac{z+t+W_{t}}{10}-\frac{6t}{250}%
\right) +\frac{6}{25},
\end{equation}

for $t\in \lbrack 0,1]$ and $z\in \mathbb{R}$.

Figure \ref{fig1} shows two realizations of the general solution in \eqref{solnex01}.
\begin{figure}[ht]
\begin{centering}
{\includegraphics[scale=0.4]{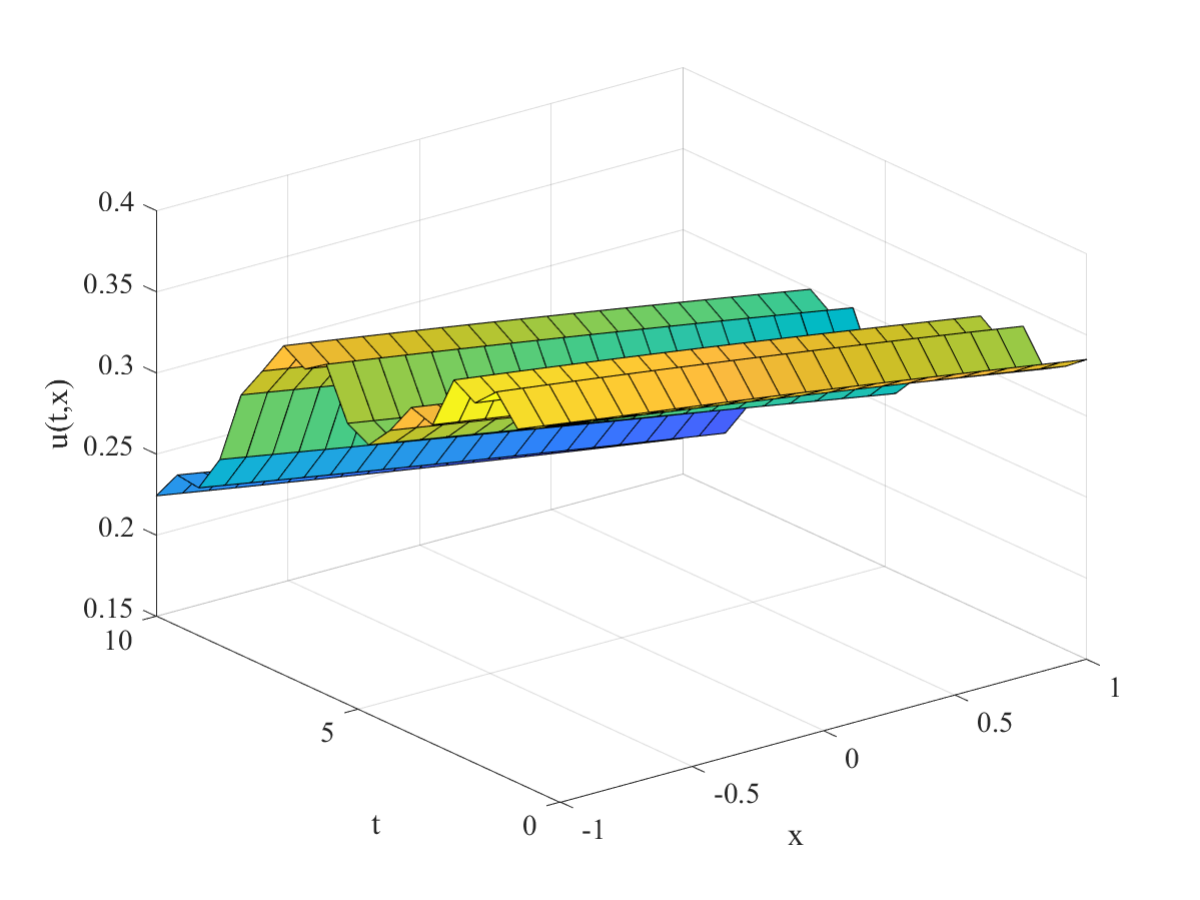}} {\includegraphics[scale=0.4]{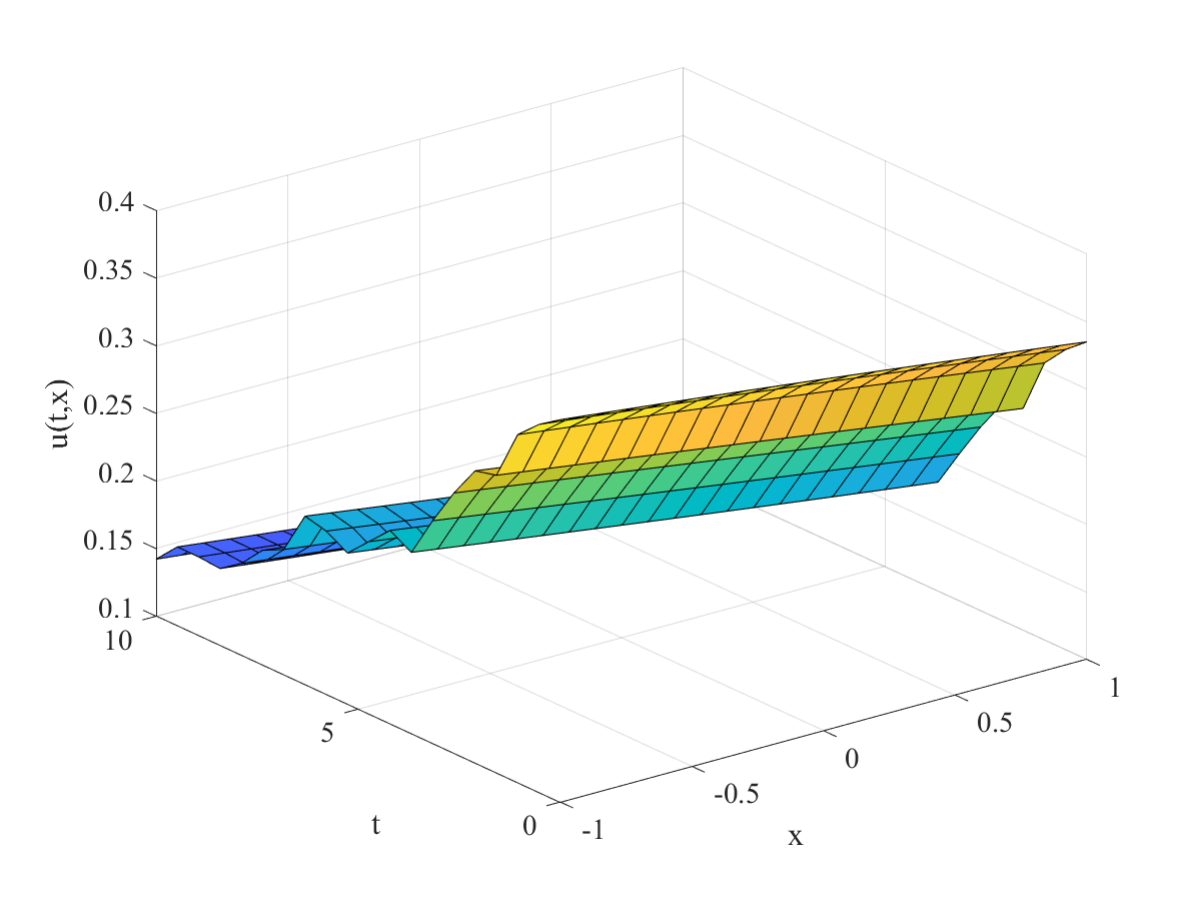}}
\par\end{centering}
\caption{Two realizations of the stochastic process in equation \eqref{solnex01}.}\label{fig1}
\end{figure}
\end{example}

\begin{example}
Consider another stochastic KdV-Burgers equation 
\begin{eqnarray}
du &=&(\partial _{zzz}u+\partial _{zz}u+u\partial _{z}u)dt+\sigma
(t)dW_{t},\quad  \label{eqex1_2} \\
u(0,z) &=&\dfrac{3}{25}\sech^{2}\left( \frac{z}{10}\right) -\frac{6}{25}\tanh
\left( \frac{z}{10}\right) +\frac{6}{25},
\end{eqnarray}

for $t\in \lbrack 0,1]$. By Proposition \ref{thm1} part (2), equation %
\eqref{eqex1_2} has a solution 
\begin{equation*}
u(t,z)=V(t,Z_{t})+\dot{Z}_{t},
\end{equation*}%
where $R(t)=1$, such that $V(t,x)$ is the solution of 
\begin{eqnarray}
\partial _{t}V &=&\partial _{xxx}V+\partial _{xx}V+V\partial _{x}V,
\label{eqex01_2} \\
\quad V(0,z) &=&\dfrac{3}{25}\sech^{2}\left( \frac{z}{10}\right) -\frac{6}{25}%
\tanh \left( \frac{z}{10}\right) +\frac{6}{25},
\end{eqnarray}%
and $Z_{t}$ is the solution of 
\begin{equation}
\ddot{Z}_{t}=\sigma (t)\dot{W}_{t},  \label{SDEex2_2}
\end{equation}%
with initial state $Z_{0}=z$ and for $t\in \lbrack 0,1]$.

Again, equation \eqref{eqex01_2} has the general solution%
\begin{equation}
V(t,z)=\dfrac{3}{25}\sech^{2}\left( \frac{z}{10}-\frac{6t}{250}\right) -\frac{%
6}{25}\tanh \left( \frac{z}{10}-\frac{6t}{250}\right) +\frac{6}{25}
\end{equation}
for $z\in \mathbb{R}$. Also, $\dot{Z}_{t}=\int_0^t \sigma(r) dW_r$ due to Lemma \ref{lemma1}. Thus, equation %
\eqref{eqex1_2} has a solution 
\begin{equation}\label{ex2_eq}
u(t,z)=\dfrac{3}{25}\sech^{2}\left( \frac{z+\int_{0}^{t}\sigma(s)dW_{s}}{10}-\frac{6t}{250}\right) -\frac{%
6}{25}\tanh \left( \frac{z+\int_{0}^{t}\sigma(s)dW_{s}}{10}-\frac{6t}{250}\right) +\frac{6}{25}+\int_0^t \sigma(r) dW_r,
\end{equation}%

If \ $\sigma (t)=1$, by Lemma \ref{lemma1}, the stochastic differential
equation \eqref{SDEex2_2} has a solution given by 
\begin{equation*}
Z_{t}=z-\int_{0}^{t}W_{s}ds
\end{equation*}%
and \ for $t\in \lbrack 0,1].$

The general solution of the stochastic KdV-Burgers equation \eqref{eqex1_2}
is given by 
\begin{equation}
u(t,z)=\dfrac{3}{25}\sech^{2}\left( \frac{z-\int_{0}^{t}W_{s}ds}{10}-\frac{6t%
}{250}\right) -\frac{6}{25}\tanh \left( \frac{z-\int_{0}^{t}W_{s}ds}{10}-%
\frac{6t}{250}\right) +\frac{6}{25}+\dot{Z}_{t}.
\end{equation}%
for $t\in \lbrack 0,1]$ and $z\in \mathbb{R}$.

If \ $\sigma (t)=t^{n}$, by Lemma \ref{lemma1}, the stochastic differential
equation \eqref{SDEex2_2} has a solution given by 
\begin{eqnarray*}
Z_{t}
&=&z-\int_{0}^{t}(t-s)s^{n}dW_{s}=z-t\int_{0}^{t}s^{n}dW_{s}+%
\int_{0}^{t}s^{n+1}dW_{s} \\
&=&z-t\left( t^{n}W_{t}-\int_{0}^{t}ns^{n-1}W_{s}ds\right)
+t^{n+1}W_{t}-\int_{0}^{t}(n+1)s^{n}W_{s}ds.
\end{eqnarray*}%
and \ for $t\in \lbrack 0,1].$

In that case, the general solution of the stochastic KdV-Burgers equation \eqref{eqex1_2}
is given by 
\begin{equation}
u(t,z)=\dfrac{3}{25}\sech^{2}\left( \frac{Z_{t}}{10}-\frac{6t}{250}\right) -%
\frac{6}{25}\tanh \left( \frac{Z_{t}}{10}-\frac{6t}{250}\right) +\frac{6}{25}%
+\dot{Z}_{t}
\end{equation}%
where $Z_{t}=z-t\left( t^{n}W_{t}-\int_{0}^{t}ns^{n-1}W_{s}ds\right)
+t^{n+1}W_{t}-\int_{0}^{t}(n+1)s^{n}W_{s}ds,$ for $t\in \lbrack 0,1]$ and $%
z\in \mathbb{R}$.
Figure \ref{fig2} shows two realizations of the general solution in \eqref{ex2_eq}.
\begin{figure}[ht]
\begin{centering}
{\includegraphics[scale=0.4]{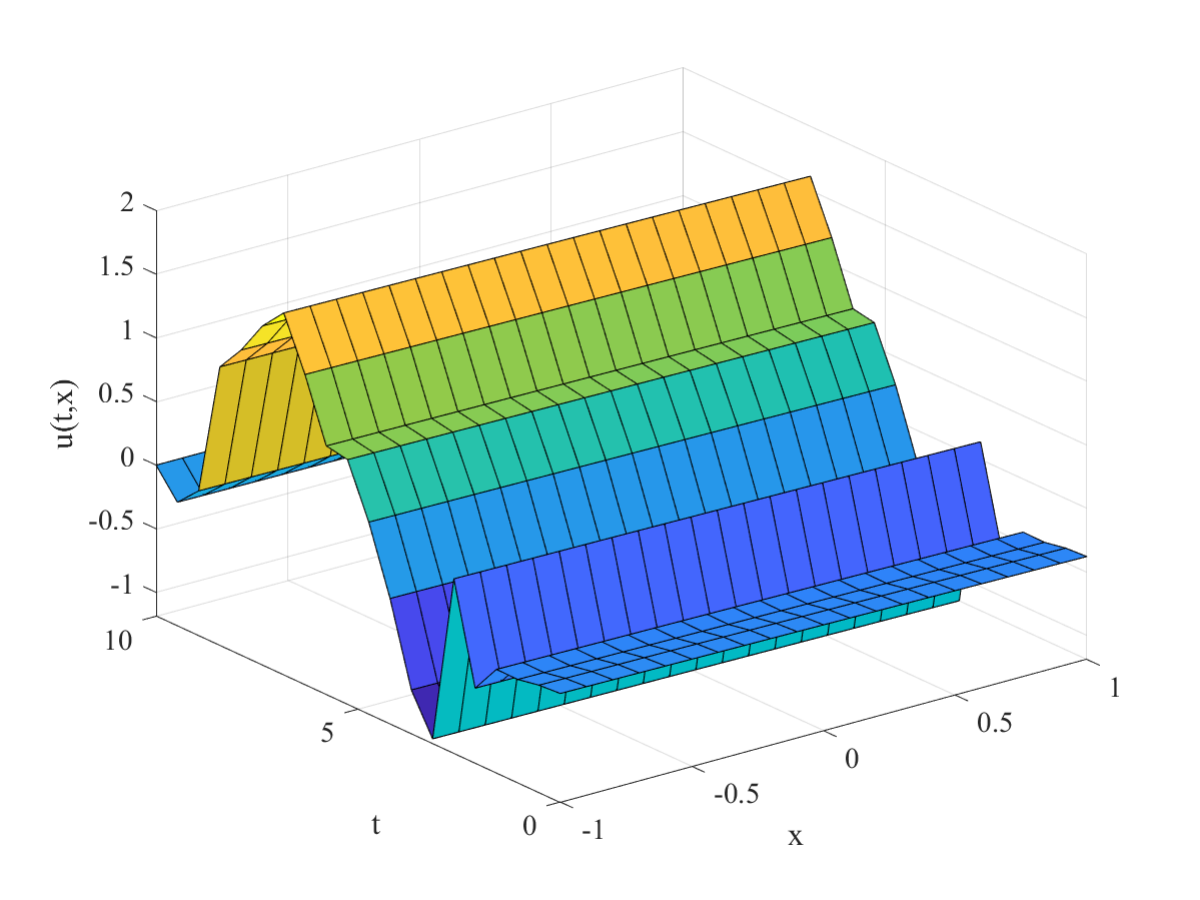}} {\includegraphics[scale=0.4]{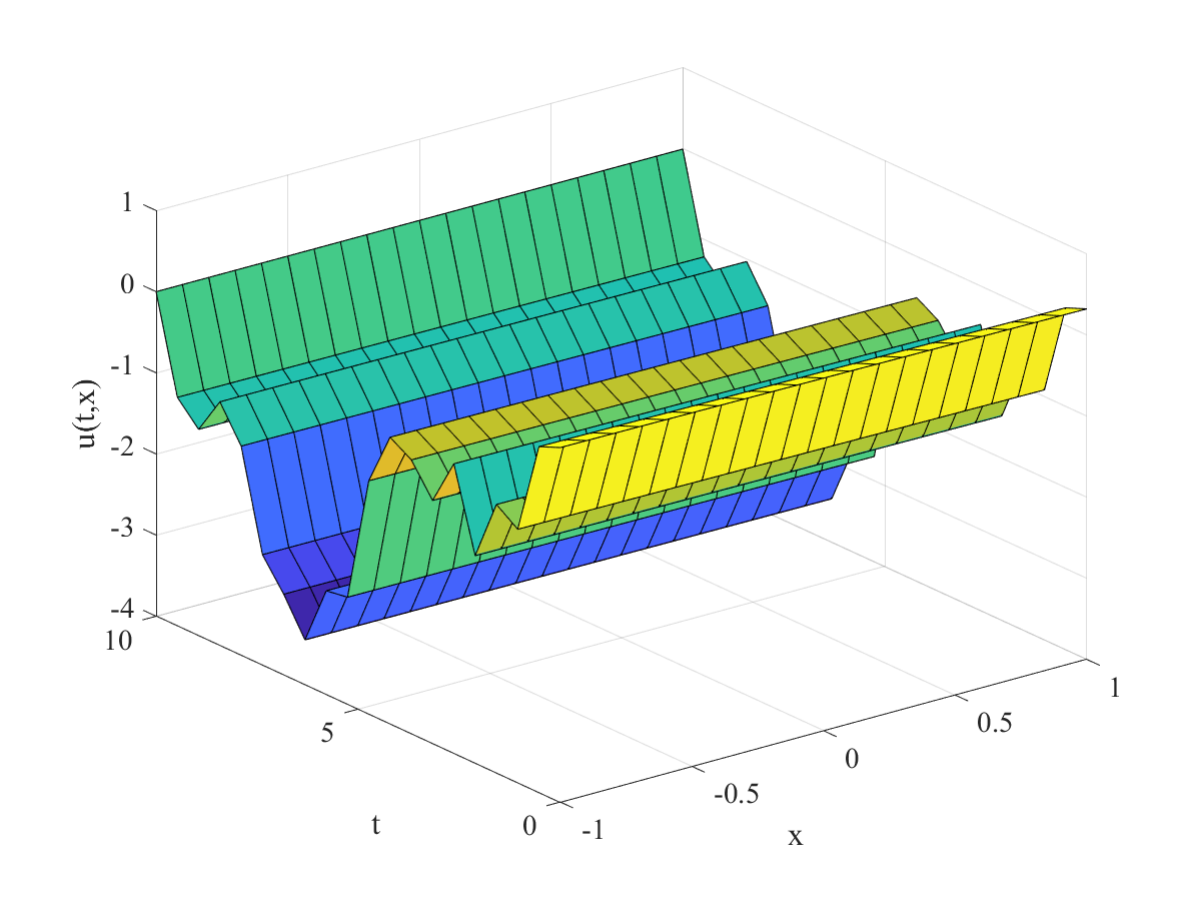}}
{\includegraphics[scale=0.4]{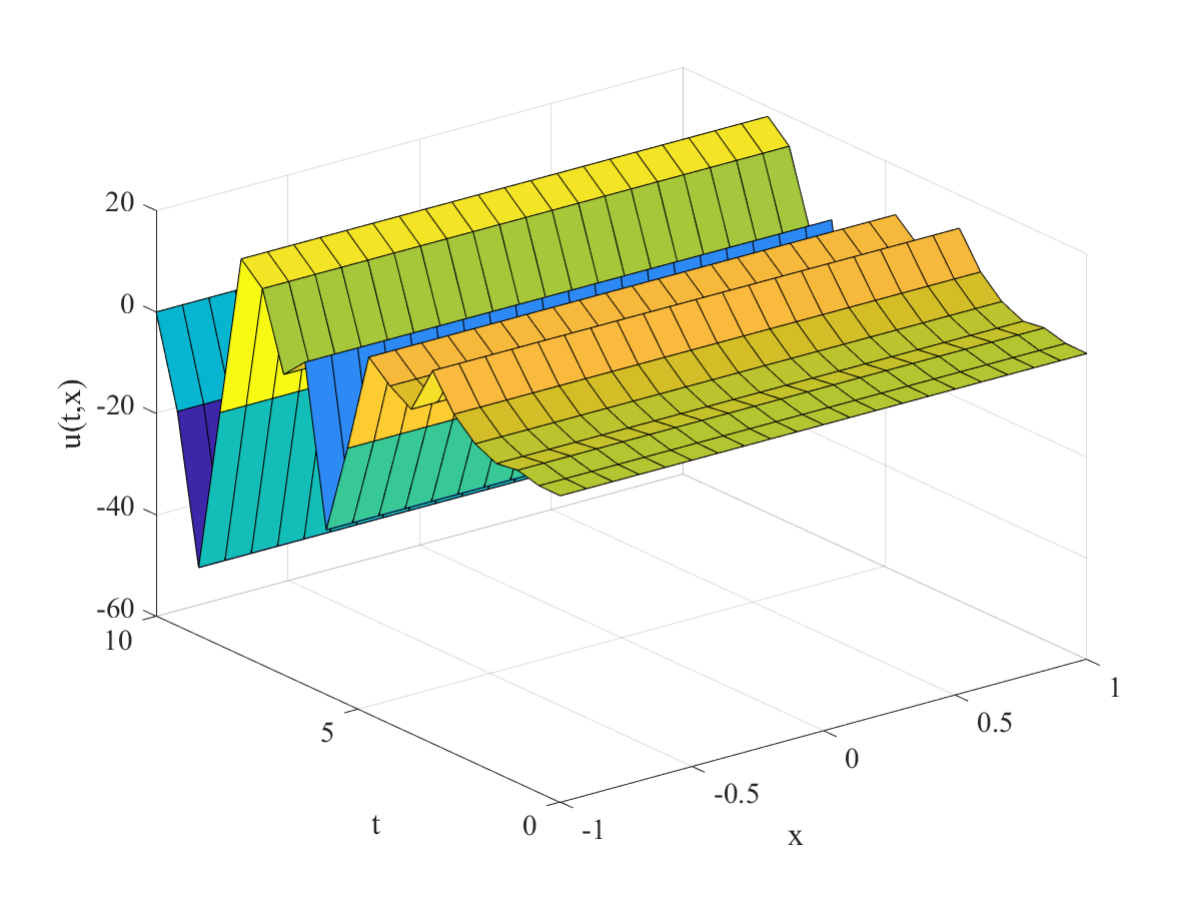}} {\includegraphics[scale=0.4]{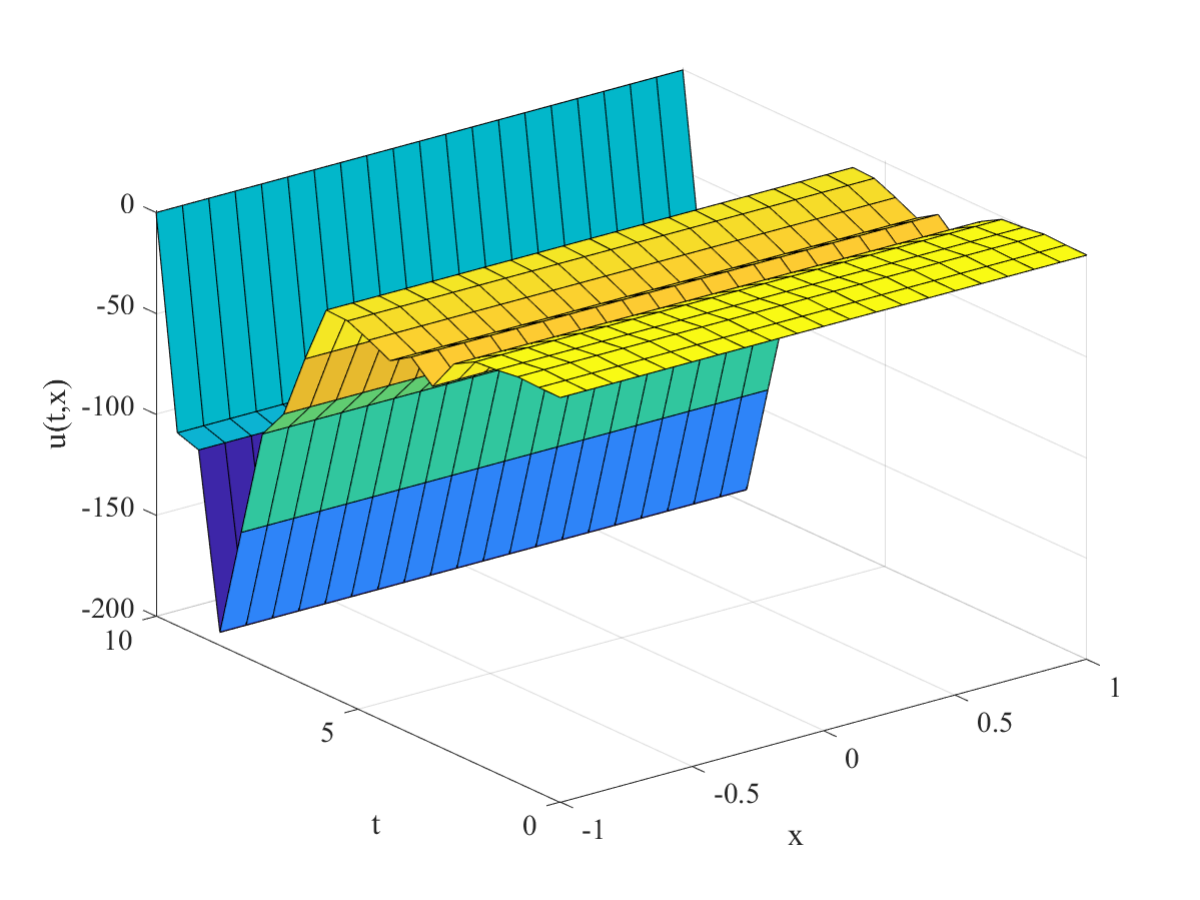}}
\par\end{centering}
\caption{Four realizations of the stochastic process in equation \eqref{ex2_eq} when $\sigma(t)=1$ (top) and $\sigma(t)=t^2$ (bottom).}\label{fig2}
\end{figure}

\end{example}

Another example for the stochastic forced term is given here.

\begin{example}
Consider another stochastic Burgers equation 
\begin{equation}  \label{eqex1_2}
du=( \exp(t) \partial_{zz}u+\exp(t)u\partial_zu)dt+ dW_t, \quad u(0,z)=%
\dfrac{2}{1+\exp(-2-z)}
\end{equation}
for $t\in[0,1]$. By Proposition \ref{thm1} part (2), equation \eqref{eqex1_2}
has a solution 
\begin{equation*}
u(t,z)= V(t,Z_t)+\dfrac{1}{\exp(t)} \dot{Z}_t,
\end{equation*}
where $R(t)=1$, such that $V(t,x)$ is the solution of 
\begin{equation}  \label{eqex01_2}
\partial_t V=\exp(t)\partial_{xx}V+\exp(t) V\partial_x V, \quad V(0,x)=%
\dfrac{2}{1+\exp(-2-x)}
\end{equation}
and $Z_t$ is the solution of 
\begin{equation}  \label{SDEex2_2}
\ddot{Z}_{t}=\dot{Z}_t+\exp(t)\dot{W}_t,
\end{equation}
with initial state $Z_{0}=z$ and for $t\in [0,1]$.

Again, equation \eqref{eqex01_2} has the general solution 
\begin{equation*}
V(t,x)=\dfrac{2}{1+\exp(-1-x-\exp(t))}
\end{equation*}
for $x\in \mathbb{R}$.

By Lemma \ref{lemma1}, the stochastic differential equation \eqref{SDEex2_2}
has a solution given by 
\begin{equation*}
Z_t=z+\exp(t)W_t-\int_0^t \exp(s) dW_s
\end{equation*}
and 
\begin{equation*}
\dot{Z}_t=\exp(t)W_t
\end{equation*}
for $t\in [0,1].$

Therefore, the general solution of the stochastic Burgers equation %
\eqref{eqex1_2} is given by 
\begin{equation}  \label{solnex01_2}
u(t,z)=W_t+\dfrac{2}{1+\exp(-1-z-\exp(t)-\exp(t)W_t+\int_0^t \exp(s) dW_s)}
\end{equation}
for $t\in[0,1]$ and $z\in \mathbb{R}$.

Figure \ref{fig4} shows two realizations of the general solution in \eqref{solnex01_2}.
\begin{figure}[ht]
\begin{centering}
{\includegraphics[scale=0.4]{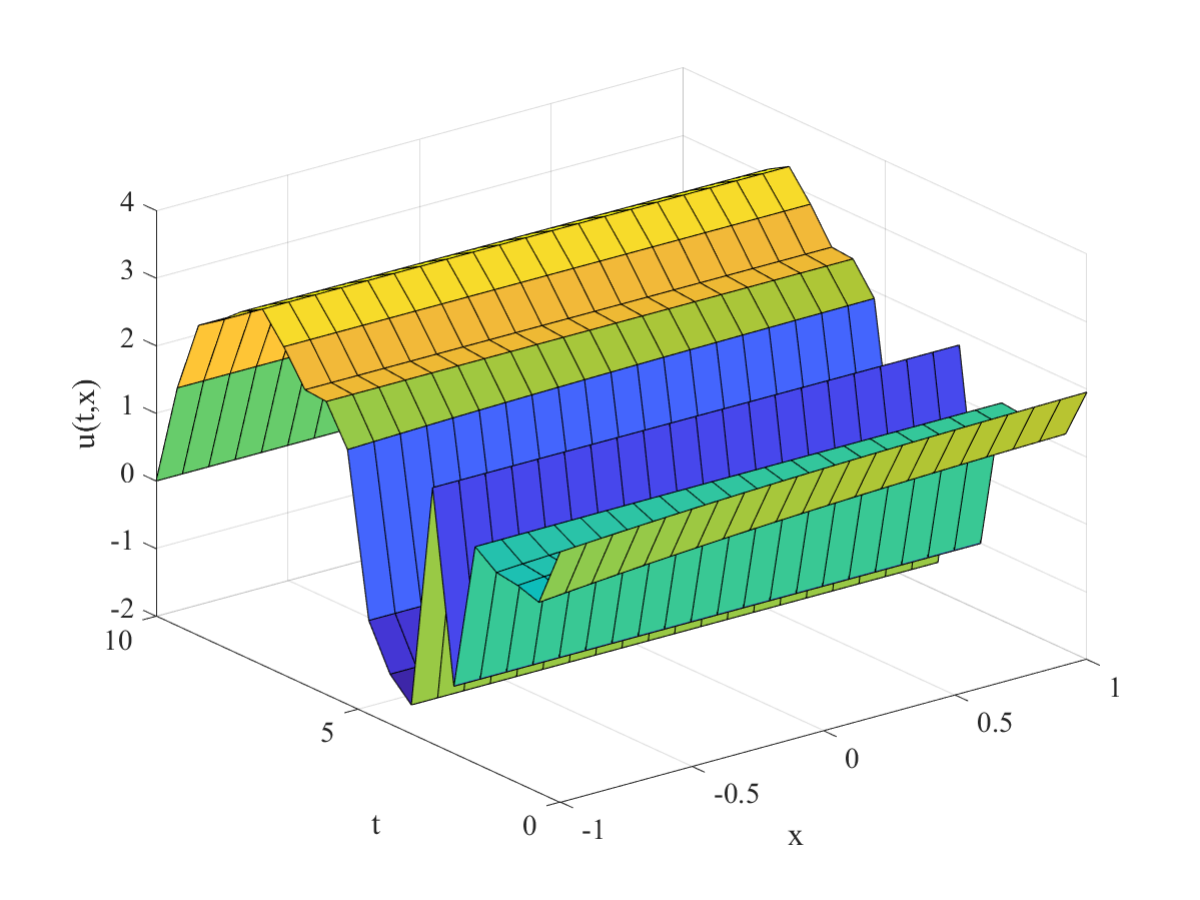}} {\includegraphics[scale=0.4]{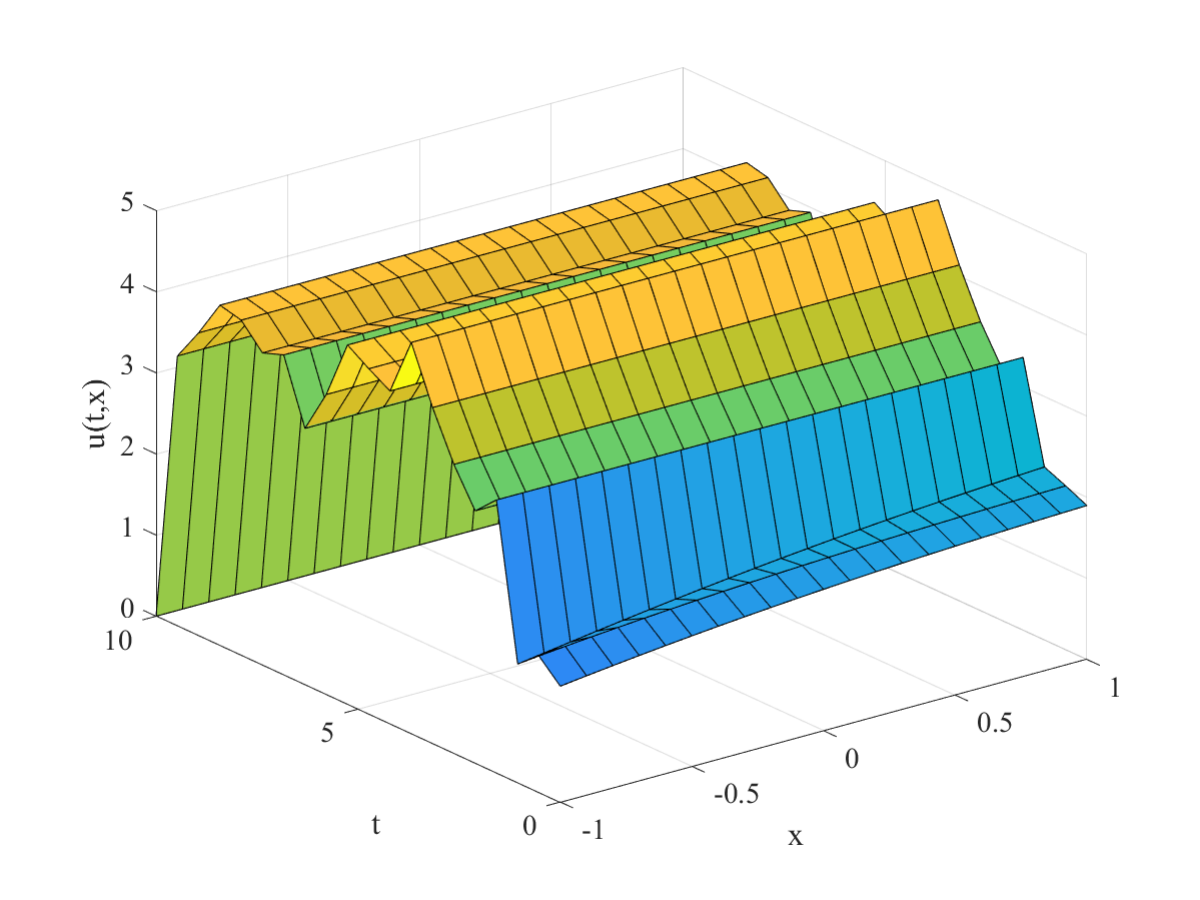}}
\par\end{centering}
\caption{Two realizations of the stochastic process in equation \eqref{solnex01_2}.}\label{fig4}
\end{figure}
\end{example}

\section{Conclusion}

In this paper, we showed using It\^o calculus that some exact solutions of
stochastic KdV-Burgers equations with several noise forms can be found using the
deterministic KdV-Burgers equations. The exact solutions were introduced and simulated.

\section*{Acknowledgments}

The authors K. Adjibi, A. Martinez, M. Mascorro and R. Sandoval where funded by the
Mathematical Association of America, award number 1652506. The author, C. Montes was funded by the National Science Foundation with award number 2150478. Faculty T. Oraby and E. Suazo were partially funded by both grants and they are the correspondent PIs. 
%The authors also thank UTRGV faculty Ms.
% Nicole Nicholson for the English proofreading and editing of this paper.

\end{document}